# Interdisciplinary Business Games on Sustainable Development: Theoretical Foundations and Prospects of Implementation[1]

*Bolshakov B.E., State University "Dubna"*

*Shamaeva E.F., State University "Dubna"*

*Popov E.B., State University "Dubna"*

*Abstract*:
The article defines the place of business games among all games in general based on the classification by F.G. Jünger; it provides critical analysis of existing business games types; it also formulates requirements and lays theoretical foundations and elements of the methodology and organization of interdisciplinary business games (IBG) on sustainable development as a special type of business games. In addition, it examines the prospects of IBG implementation in higher education for sustainable development, using information technology and computer resources. The work has been written as a part of RFBR project №12-06-00286-a.

Business game is a very special type of activity and learning tool. It combines rational decisions and irrational excitement. With this in mind, let us consider for a start, which place business games take up among the variety of games known to man.

Some thoughts on this issue have already been given in [7], so we will not dwell on them again, and turn to the work of Friedrich Georg Jünger "Games. The Key to Their Meaning" [14]. Jünger writes that it is impossible to talk about the origin of the "game in general", "the game as it is" — you can only talk about the origin of specific games [14, p. 42]. This emergence of games is a process that has not remained somewhere in the past, in the early history of mankind. This process takes place at all times, it is always runs along with the game: "Games appear as they always appeared" [14, p .43].

Jünger subdivides games into three kinds — depending on the basis on which they function [14, p. 39-44]. He identifies:
- games based on lucky chance;
- games based on mastery;
- games based on imitation — anticipatory (*pre-imitation*) or retroactive (*post-imitation*).

Let us consider the last, third kind of games. What is it like? "Post-imitation, which binds itself to events occurred in the past, is an image of the past movement presented by the same or a similar movement. Pre-imitation which connects itself with what happens in the future, is an image of the future movement presented by the same or a similar movement". Jünger illustrates these concepts with the example of a child learning the native language (as a child repeats the words he or she heard, and "makes up" new ones — collects sounds into syllables and syllables into words) [14, p. 85]. Going directly to the games, he cites the example of a girl playing with a doll: this girl mimics the behavior of a mother caring for her child (she mimics this behavior, but

---

does not copy certain movements mechanically) — i.e. "…does something she will do in the future, no longer playing" [14, p. 88]. And here appears the inextricable link between pre- and post-imitation: "[The power of imitation] stems from the continuity of space and time <...> our own movement, the movement of the identical person reflects this continuity, following it. Therefore, imitation, which subsequently or preliminary displays something is — if you look at it from the perspective of time and space — the return and repetition of the same or similar. <...> And the fact that the scope of role model goes far beyond the bodily movements is indicated by our memories.

If we are aware of the fact that everything stems from the imitation of the continuity of time and space, we understand that a standalone post-imitation can be understood only as a post-imitation without pre-imitation, that is, as the end of continuity, or discontinuity. The same can be said about the pre-imitation without post-imitation» [14, p. 86].

We can finally distinguish the key point: "The hallmark of pre-imitative and post-imitative games is that they reflect in the form of a game a certain behavior that is not a game" [14, p. 88]. This can be said about business games — they certainly belong to this kind of games[2]. We give here a quote directly related to business games which illustrates this: "Once they (the players — *auth.*) have realized that they have new goals, <...> as soon as they got it and realized the fact that these goals are not translated into the problem, that there is no mode of action corresponding to these goals — they will naturally turn to the same mode of action to the same procedures, begin to break them, transform <...> and at that moment they catch themselves and say, "what is this nonsense I do and say?". And you encourage them and say: "We're playing a game here". And in the game you can do anything. You can take a chair and say that you are riding a train" [13, p. 71].

There are different classifications of business games. We will consider some of them.

In [12], a *chronological* classification of business games is given — on the basis of certain stages in the development of social gaming technology:
- the first generation: closed type (strict games) — "... the main result was the mastery of specific skills in a particular situation. <...> Freedom was limited to a freedom to seek solutions. But this search was carried out under strictly-defined conditions: time, place and actors" [12, p. 61-62], these games were "strict in terms of rules and closed in terms of freedom"[3] [12, p. 62]; their examples include the first business games in the USSR (M.M. Birshtein, 1932-38) and the USA (J. von Neumann, 1920s., O. Morgenstern, 1940s.);
- second generation: open type (free games) — the open character of this generation manifests itself in everything: there are no strict guidelines for choosing the topic and selecting the participants, there's also a wide variety of game organization methods; the main objective of these games — the formation of the new way of thinking among participants [12, p. 62-63]; these include, first of all, organizational activity games (OAG) by G.P. Shchedrovitsky (1970s.); the derivatives of OAG developed by former members of the currently non-functioning Moscow Methodological Circle usually don't belong exclusively to this type;
- third generation: open type (regulated games) — the main difference from the second generation is the careful selection of the participants (they must have direct relevance to the game topic by virtue of their professional activities — such as customers and executors of a particular project), and the restrictions imposed on the game methodologist's actions: he or she should take into account the consequences of the games — as above-mentioned OAG literally "pulled the rug from under the feet" of players; not everyone could deal with such aggressive influence, and some participants and often left the game long before its end; one of the earliest examples of the third generation games is an innovative game by V.S. Dudchenko (mid-1980s). [12, p. 62-63].

---

[2] The authors define a business game according to [2, 7].

[3] A typical example of this type of games is described in [4, p. 204-205]: "The audience is given a network matrix in which the stages of decision-making are superimposed on the time grid with vertical transfer of all business leaders. The algorithm turns into a roadmap with its critical path. Arrows go from block to block — you can immediately see, to whom and in what timeframe it is necessary to perform a particular task — who should act immediately, and who has more time".

Currently the most widely used business games in education belong to the first and the third generation, or combine features of two or even all three generations.

In [10] a classification based on *gaming purposes and their scope* is proposed:

- educational business games — are used as the name implies, in the learning process of pupils, students and workers for modeling problematic situations that they may encounter in their future (or present — in case of employees) activities;
- validation games — used to assess the qualifications of employees, or to identify the participants' knowledge on a particular subject (these also include job interviews);
- research games — used to test new methods of improving the management and control, i.e. in cases where the consequences of a failed field experiment can be fatal;
- industrial games — a type of decision-making methods, built into the actual management process.

Furthermore, mixed types of games are possible: for example, one and the same game can belong to educational and research type, etc.

There are other kinds of classifications, but to identify the strengths and weaknesses of existing approaches to development, organization and carrying out of business games it will be more convenient to use the above-mentioned chronological classification. The analysis will pay special attention to the "technologization" possibility for the games of different generations (Table 1).

**Table 1. Analysis of approaches to the development, organization and carrying out of business games.**

|  | Advantages | Disadvantages |
|---|---|---|
| **First generation games** | - the relative simplicity of game development and entry into the game — provided that the participant has adequate professional competence<br>- clear and specific terms, goals and objectives of the game and modelled situations | - the applicability of this type is limited to the narrow professional tasks<br>- limited freedom of action does not allow participants to fully deploy the creative process |
| **Second generation games** | - broadest coverage of simulated situations (no limitations on the scope and topic) in conjunction with the "low threshold" of entering the game: participants are not required to be professionals in the field of activity, which is modeled in the game<br>- formation of the new way of thinking among participants | - rather aggressive influence on the participants in order to build up their new way of thinking through the break-up of the old<br>- due to the specifics of these games their "technologization" is difficult or impossible |
| **Third generation games** | - these games overcome the lack of the second generation associated with the aggressiveness of the impact on the participants<br>- more rigorous organization of the game opens up opportunities for its "technologization", while keeping the creative component intact | - less broad thematic scope than that of the second generation of games<br>- "higher threshold" of entering the game: participants should be professionals in the area of activity |

Strengths and weaknesses revealed above allow us to formulate the following requirements for the business games:
- from the perspective of sustainable development — the requirements of *interdisciplinarity* and *measurability* [1];
- from the perspective of modern education —*"technologization"* claim [3, 7, 16];
- a general requirement — *focus on development*.

Staff of the Department of Sustainable Innovative Development (SID) of the State University "Dubna" and the International Scientific School of Sustainable Development named after P.G. Kuznetsov creates interdisciplinary business games (IBG) on sustainable development satisfying these requirements. Developed IBGs are being successfully implemented in the educational process (see [2] and future publications on this topic in the next issues of electronic scientific magazines "Sustainable Innovative Development: Design and Management" and "Sustainable Development: Science and Practice").

IBGs are called so due to the *interdisciplinary nature* of sustainable development itself, covering environmental, economic, and social sphere, since sustainable development as a branch of science is a synthesis of the natural, social and humanitarian sciences, which overcomes the existing "dimensional gaps" between these sciences [1, 5]. Thus, the general of IBG is to convert the existing "dimensional gaps" into a driving force of development through: *the transfer of knowledge and the generation of new knowledge, or innovations* (the first sub-goal) and *the formation of the new way of thinking* among the participants of the game (the second sub-goal). The first sub-goal is the closest, it is achieved directly in the game, and the second sub-goal is remote, as the game "starts" the formation of the new way of thinking.

Let us consider these two sub-goals starting with the first one: IBG can be a means of knowledge transfer, i.e. possess traits of a reproductive method of teaching, but it is not a lecture. Note: "Knowledge, fixed memory is preserved only to the extent that it is supported by thinking. Knowledge needs the process of thinking to reproduce itself; therefore, the latter should always be present". [1]. "The thinking and knowledge in general are inseparable. And where the mind acts as the discoverer of new knowledge, it implies at the same time the use of existing knowledge. This is the basis for the theory that thinking is the process of functioning and actualization of knowledge. This <...> is not correct, if by this we mean that thinking comes down to only functioning of knowledge. This is correct, if we understand that the functioning of knowledge is involved in the process of thinking" [8, p. 53]. Knowledge obtained (generated) during the IBG is assessed in terms of its quantity — *the amount of new knowledge* — and quality — on the basis of its *contribution to the development of the control object* in a specific simulated situations [7].

Let us move on to the second sub-goal: *thinking of sustainable innovative development* is the way of thinking necessary for the transition to sustainable innovative development at the expense of personal contribution (creativity) of each individual carrier of this way of thinking. Under the creativity here we mean the process of *ideas* generation — those that contribute to the development of the control object, and their subsequent implementation. The formation of this way of thinking is the result of *psychological adjustment* during the game, which begins with the creative process and is required to maintain it.

Now we will briefly outline the way IBG is conducted. Generally any IBG can be divided into three phases. During *the first phase*, we stimulate the activity of game participants, which then leads to the first conflict; participants begin to understand their roles, identify the contradictions between their goals, and that gives rise to the first conflict and the transition to the second phase. During *the second phase*, the first conflict leads to the necessity of self-organization, association, which generates the second conflict; self-organization of the participants in this phase is reduced, ultimately, to the ability to negotiate; after reaching an agreement we reveal its shortcomings, and this creates a second conflict and the transition to the third phase. During *the third phase*, the second conflict is removed in the form of self-development of game participants: they think of options to improve the agreement to which they had come, or come to a new agreement, free of revealed deficiencies.

All IBGs can be fairly divided into *highly regulated* and *weakly regulated* on the basis of the rules of the game[4]. In the first case, the choice of possible actions for the game participants is limited much more than in the second, which is evident from the titles of both.

The experience of IBGs conduction includes the following most notable games:

- *"Money: yesterday, today, tomorrow"* — IBG devoted to the role of money as a medium of exchange based on the teachings of Karl Marx on the socio-economic formations; example of a weakly regulated games (successfully carried out several times within the framework of a permanent seminar of the Department of SID, as well as with the students of distance learning at the same department [2]);
- *"River Banks Facilities"* — IBG with an environmental awareness topic, which simulates the interaction of facilities placed on the banks of the same river, which take the water from it for their industrial purposes; notable for the first introduction of random events as part of the gameplay; example of a strongly regulated games (successfully carried out in November 2014 with the first-year master students of the Department of SID);
- *"Fishermen"* — IBG devoted to the development of mutually beneficial conditions of the shared resource use (in this case — the fish in the coastal waters zone) on the basis of studies by E. Ostrom [6, 15] (scheduled for the spring semester of 2015 for full-time students of the Department of SID);
- *"Mankind's Transition to Autotrophy"* — IBG devoted to the opportunities and possible ways of mankind's transition to autotrophy according to V.I. Vernadsky (scheduled within the framework of scientific and educational activities of the Department of SID in April 2015).

We now turn to the above-mentioned requirement of "technologization" because we have considered the other requirements more or less exhaustively. "Technologization" means among other features the involvement of certain technical means to empower the implementation of IBG. Here we will certainly encounter information technology and computer systems.

There are several options for "computerization" of business games:

- *fully integrated solutions* — in this case, the business game is a complete software product, a certain electronic "environment" in which players act; an example of this type is the well-known interactive computer business game "Nixdorf Delta" [9], which is a simulation model of the company development for several years.

The unquestionable advantages of this type of solutions include their "finished form" — all game actions take place within a single software product, it allows participants to fully focus on the gameplay, gives them the opportunity to "dive into the game"; it also does not require the study of a whole set of software tools. But there are also disadvantages: low flexibility (limited number of program options embedded in game situations and, in addition, they are of the same type) and focus on the interaction between man and computer — interpersonal communication here go by the wayside, the participants play more with a computer than with each other.

- *modular solutions* — in this case we use a complex of certain software products (i.e. modules), most of which are not specifically designed for use in a business game; however, there is some basic program — so to speak, the "body" of the game, which the player is free to add other modules to as auxiliary.

This type is undoubtedly more flexible in comparison with the first. However, players face the difficulties associated with the use of a number of various software products. Of course, the study of these is beneficial to participants of the game, but at the same time, it either slows down the gameplay, or makes preparations for the game too long. Furthermore, there may be compatibility issues, particularly if each of the programs stores data in its own specific file format.

---
[4] The concept of "rules of the game" is introduced and defined according to [7].

- *instrumental and auxiliary solutions* — in this case, there are no restrictions on the use of software, but they are complementary; a central role here is played by the interaction between the players themselves.

The most primitive version of such a solution is the use of Skype for establishing communication between players located geographically far apart during the "classic" business game. However, in general, the use of a special software for the simulation of various processes modelled in the game is expected — so the players could clearly observe the consequences of their decisions, to support their points of view or to resolve the conflict. The simplest example is charting in Microsoft Excel to monitor the dynamics of the process for the given boundary conditions.

We see instrumental and auxiliary solutions as optimal for IBGs, as for them, just like in the "classic" non-computer games business, the focus is on the interaction of participants with each other — it is essential for achieving the above-mentioned sub-goals of IBG. Of course, these solutions are not without certain disadvantages: firstly, it is complicated to implement random events into the game (based on the experience of the non-computer IBGs we can be argue that random events heat interest of participants throughout the gameplay, motivate them to be more active); secondly, the requirements for the players are quite high, as those in the case with modular solutions — it may take a long time to master various software.

Organization and carrying out computer-based IBGs by means of instrumental and auxiliary solutions is planned in 2015 on the base of the Institute of System Analysis and Management of the State University "Dubna". At the same time non-computer interdisciplinary business games on sustainable development, created on the basis of theoretical and methodological foundations of the organization and conduction of IBG developed by the authors, have been successfully used to model the situations in the process of training specialists for sustainable innovative development of the country, and can be used to make organizational and managerial decisions based on the principles of sustainable development and taking into account its requirements, in the state, public and business structures in the presence of conflict or uncertainty.

*Information about the authors*:

**Bolshakov Boris Evgenievich** — doctor of technical sciences, professor, academician of the Russian Academy of Natural Sciences, Head of the Department of Sustainable Innovative Development at the Institute of System Analysis and Management of the International University of Nature, Society and Man "Dubna", Head of the International Scientific School of Sustainable Development named after P.G. Kuznetsov.
Address: 141982, Moscow Region, Dubna, Universitetskaya str., 19/1, office 428.
Tel.: (496) 216-61-09
E-mail: bb@uni-dubna.ru

**Shamaeva Ekaterina Fiodorovna** — candidate of technical sciences, associate professor of the Department of Sustainable Innovative Development at the Institute of System Analysis and Management of the International University of Nature, Society and Man "Dubna".
Address: 141982, Moscow Region, Dubna, Universitetskaya str., 19/1, office 428.
Tel.: (496) 216-61-09
E-mail: school@yrazvitie.ru

**Popov Eugene Borisovich** — assistant of the Department of Sustainable Innovative Development at the Institute of System Analysis and Management of the International University of Nature, Society and Man "Dubna", leading interdisciplinary business games developer.
Address: 141982, Moscow Region, Dubna, Universitetskaya str., 19/1, office 428.
Tel.: (496) 216-61-09
E-mail: mc.insekt@gmail.com